\documentclass[12pt,a4paper]{article}

\usepackage{amsmath,amssymb}
\usepackage[bookmarks=true]{hyperref}
\usepackage[nosort]{cite}
\usepackage{bbm}
\usepackage[bulletsep]{collect}
\def\gfxon{\usepackage[final]{graphicx}}

\gfxon

\makeatletter
\let\old@startsection=\@startsection
\renewcommand{\@startsection}[6]{\old@startsection{#1}{#2}{#3}{#4}{#5}{#6\mathversion{bold}}}
\makeatother

\def\eeq{\end{eqnarray}}

\def\p{\partial}

\def\de{\partial}

\def\=:{=\hspace{-.7em}\raisebox{1.1ex}{.}\hspace{.1em}\raisebox{-0.2ex}{.} }

\newcommand{\beqn}{\begin{eqnarray}}
\newcommand{\eeqn}{\end{eqnarray}}
\newcommand {\beq}{\begin{eqnarray}}
\def\eeq{\end{eqnarray}}
\newcommand {\eeqq}{\end{eqnarray}}
\newcommand {\non}{\nonumber\\}

\newcommand {\tr}{{\rm tr}\,}
\newcommand {\Tr}{{\rm Tr}\,}

\makeatletter \@addtoreset{equation}{section} \makeatother

\makeatletter
\let\old@makecaption=\@makecaption
\def\@makecaption{\small\old@makecaption}
\makeatother

\makeatletter
\def\mr@ignsp#1 {\ifx\:#1\@empty\else #1\expandafter\mr@ignsp\fi}%
\newcommand{\multiref}[1]{\begingroup
\xdef\mr@no@sparg{\expandafter\mr@ignsp#1 \: }%
\def\mr@comma{}%
\@for\mr@refs:=\mr@no@sparg\do{\mr@comma\def\mr@comma{,}\ref{\mr@refs}}%
\endgroup}
\makeatother

\ifx\href\asklfhas\newcommand{\href}[2]{#2}\fi

\begin{document}

\begin{flushright}\footnotesize
\texttt{IFUP-TH/2012-12} \\
\texttt{}
\vspace{0.5cm}
\end{flushright}
\vspace{0.3cm}

\renewcommand{\thefootnote}{\arabic{footnote}}
\setcounter{footnote}{0}
\begin{center}%
{\Large\textbf{\mathversion{bold}
Supersymmetry Breaking \\  on \\ Gauged  Non-Abelian Vortices}
\par}

\vspace{1cm}%

\textsc{Kenichi Konishi$^{1}$, Muneto Nitta$^{2}$,   Walter Vinci$^{1}$}

\vspace{10mm}
$^{1}$\textit{Department of Physics, ``Enrico Fermi`''\\ Pisa University, 
Largo Pontecorvo,3, 56127, Pisa, Italy }
\\
\textit{INFN, Sezione di Pisa,  
Largo Pontecorvo,3, 56127, Pisa, Italy}
\\
\vspace{2mm}
$^{2}$\textit{Department of Physics, and Research and Education Center 
for Natural Sciences,\\ Keio University, 4-1-1 Hiyoshi, Yokohama, 
Kanagawa 223-8521, Japan}
\\
\vspace{.3cm}

\vspace{7mm}

\thispagestyle{empty}

\texttt{konishi(at)df.unipi.it}\\
\texttt{nitta(at)phys-h.keio.ac.jp}\\
\texttt{walter.vinci(at)pi.infn.it} 

\par\vspace{1cm}

\vfill

\textbf{Abstract}\vspace{5mm}

\begin{minipage}{12.7cm}

There is  a great number of systems characterized by a completely broken gauge symmetry, but 
with an unbroken global color-flavor diagonal symmetry, \emph{i.e.}, systems in the so-called color-flavor locked phase. If the gauge symmetry breaking supports 
vortices, the latter develop non-Abelian orientational zero-modes and  become non-Abelian vortices, a subject of intense study in the last several years. 
In this paper we consider the effects of weakly gauging the full exact global flavor symmetry in such systems, deriving an effective description of the light excitations in the presence of a vortex. Surprising consequences are shown to follow.
The fluctuations of the vortex orientational modes  get diffused  to bulk modes through tunneling processes.  When the model is embedded in an ${\cal N}=2$ supersymmetric theory, the vortex is still 1/2 BPS saturated, but the vortex effective action breaks supersymmetry spontaneously.  

\end{minipage}

\vspace{3mm}

\vspace*{\fill}

\end{center}

\newpage

\section{Introduction}

Since their discovery~\cite{Auzzi:2003fs,Hanany:2003hp,Hanany:2004ea,Shifman:2004dr}, non-Abelian vortices have provided a powerful tool to study various aspects of the non-perturbative dynamics of non-Abelian gauge theories and non-linear sigma models. Non-Abelian vortices arise in the ``color-flavor'' locked phase of a non-Abelian gauge theory, where the vacuum preserves a non-Abelian global symmetry which is a diagonal combination of gauge and flavor symmetries. A non-Abelian vortex breaks further this residual symmetry, and non-Abelian degrees of freedom are generated as Nambu-Goldstone (NG)  modes confined on the vortex worldsheet. The most typical example of theory with this type of solitons is a $U(N)$ gauge theory with $N$ scalars (flavors) in the fundamental representation and a color-flavor symmetric scalar potential (and masses) that puts the theory on the Higgs phase:
\begin{equation}
U(N)_{C}\times SU(N)_{F} \stackrel{vac}{\longrightarrow} SU(N)_{C+F} \,.\nonumber \\
\end{equation}
A vortex solution would then break the residual symmetry as follows:
\begin{equation}
 SU(N)_{C+F} \stackrel{vortex}{\longrightarrow} SU(N-1)_{C+F}\times U(1)_{C+F} \,.	\nonumber \\
\end{equation}
In this case the vortex develops zero-modes which are described by a non-linear sigma model on a $\mathbb CP^{N-1}$ target space:
\begin{equation}
\mathcal M=\mathbb CP^{N-1}=\frac{SU(N)_{C+F}}{ SU(N-1)_{C+F}\times U(1)_{C+F}}.\nonumber \\
\end{equation}
The zero-modes above describe internal degrees of freedom of the vortex, or, in other words, the orientation of its non-Abelian flux into the non-Abelian gauge group. The mixing with flavor symmetries prevent the possibility that these zero-modes can be gauged away. They are thus physical and generalize the translational zero-mode of the usual Abrikosov-Nielsen-Olesen string~\cite{Abrikosov:1956sx,Nielsen:1973cs}. After their discovery, non-Abelian vortices have been constructed and studied in a wide class of non-Abelian gauge theories, with arbitrary number of flavors~\cite{Eto:2007yv,Shifman:2006kd,Shifman:2011xc,Auzzi:2008wm}, were semi-local vortices arise~\cite{Vachaspati:1991dz,Hindmarsh:1991jq,Hindmarsh:1992yy,Preskill:1992bf,Achucarro:1999it}, and different gauge groups~\cite{Eto:2008yi,Eto:2009bg}.

While one can consider non-Abelian vortices in generic non-Abelian theories, provided we also include non-Abelian flavor symmetries, the most spectacular results have been obtained in theories with extended $\mathcal N=2$ supersymmetry, or in the case where  $\mathcal N=2$ is softly broken to  $\mathcal N=1$ by mass terms for the adjoint fields. In this case vortices are 1/2 ``BPS''-saturated~\cite{Bogomolny:1975de,Prasad:1975kr}, which means  they preserve half of the supercharges present in the bulk theory and the effective theory  is then a  $\mathbb CP^{N-1}$  non-linear sigma model with $\mathcal N= (2,2)$ extended supersymmetry in 2 dimensions. In these systems, non-Abelian vortices provide a physical exlpanation~\cite{Hanany:2004ea,Shifman:2004dr,Tong:2003pz} for the long-known correspondence between the mass spectrum of $\mathcal N=2$ gauge theories in four dimensions and $\mathcal N= (2,2)$ sigma models in two dimensions~\cite{Dorey:1999zk,Dorey:1998yh}. These results rely on the presence of exact formulas for the mass of BPS saturated states, as holomorphic functions of the couplings~\cite{Seiberg:1994rs,Seiberg:1994aj,Witten:1993yc}. In this picture, the two-dimensional states are seen as four-dimensional states ``confined'' on the vortex worldsheet. Another crucial  result has been to recognize the importance of the monopole-vortex complex, in theories with $\mathcal N=1$ soft mass terms, to study the properties of non-Abelian confinement~\cite{Auzzi:2003fs,Auzzi:2003em,Auzzi:2004if,Eto:2006dx}, and to identify the dual quarks present in some quantum vacua with residual non-Abelian symmetry ($r$-vacua) as light non-Abelian monopoles~\cite{Carlino:2000ff,Carlino:2000uk}.

Since the very existence of non-Abelian and semi-local vortices is due to the rich interplay between gauge and flavor symmetries, it is in our opinion very interesting to investigate   how the classical and quantum properties of these solutions get modified when some of the flavor symmetries are gauged.

 This is the main purpose of our paper. 
 
 Gauge theories with partially or fully gauged flavor symmetries have been extensively considered, for example, in the context of quiver theories~\cite{Douglas:1996sw}. In a more realistic setup, the gauging of a $U(1)$ flavor symmetry means taking into account of electromagnetic interactions in the color-flavor-locked phase of QCD at high density and low temperatures~\cite{Alford:2007xm}, where non-Abelian vortices have been shown to exist~\cite{Balachandran:2005ev,Nakano:2007dr,Nakano:2008dc,Eto:2009kg,Eto:2009bh,Eto:2009tr}. A related work of some of the authors explicitly studied the effects of the gauging of flavor symmetries on supersymmetric non-linear sigma models, showing that the moduli space metric of solitons can be deformed, and even regularized in the case of lumps and semi-local vortices~\cite{Nitta:2011um}. The present work can be regarded as a continuation of that project, extending it to non-Abelian gauge theories and non-Abelian vortices.

In Section 2 we briefly review the construction of non-Abelian vortices in theories with $SU(N)$  flavor symmetries. In Section 3 we consider the effects of weakly gauging the whole flavor symmetry, and  derive the effective action on the vortex worldsheet. We find that the vector-like ``color-flavor''  symmetry is now realized as an unbroken gauge symmetry in the bulk.  Accordingly the worldsheet effective action takes the form of a gauged non-linear sigma model.  The unbroken  gauge symmetry is broken further along the vortex core, giving mass to  some of these gauge bosons, in a sort of vortex-induced Higgs mechanism.  It might appear at first sight  that our vortex is an elegant non-Abelian generalization of Witten's superconducting string~\cite{Witten:1984eb}.  Actually, as we discuss in Section 4,  these light gauge bosons are unstable against decay into the bulk as massless gauge bosons, due to the typical non-Abelian cubic or higher gauge boson interactions, absent in the $U(1)$ model considered by Witten.  Finally, in Section 5 we discuss the BPS limit and the embedding of our systems into a supersymmetric theory.  We find, somewhat unexpectedly, that even though our BPS vortex (classical configuration) respects half of the four-dimensional supersymmetries, the effective action on the vortex worldsheet  appears  necessarily to break supersymmetry.
We summarize our results in Section 6.  In Appendix we give some more details of the derivation of the vortex effective action.  

\section{Non-Abelian Vortices}

Let us briefly review the construction of non-Abelian vortices in non-Abelian gauge theories. The simplest case a non-Abelian vortex arises is an $SU(2)\times U(1)$ gauge theory with two scalar fields in the fundamental representation and a $SU(2)$ flavor symmetry~\cite{Auzzi:2003fs,Hanany:2003hp,Hanany:2004ea,Shifman:2004dr,Shifman:2007ce,Eto:2006pg,Tong:2005un,Tong:2008qd}: 
\begin{eqnarray}
\mathcal L    &= &  \Tr\left[   \frac1{2 }F_{\mu\nu}^{2}+ |\mathcal D_{\mu} Q|^{2} +\lambda_{2}(\bar Q  Q)^{2} -m^{2} \bar Q  Q  \right]+\lambda_{1}(\Tr[\bar Q Q])^{2}+\frac{ m^{4}}{2(2\lambda_{1}+\lambda_{2})}\,,\nonumber \\
  \label{eq:act}
\end{eqnarray}
where the $Q$ are the scalar fields written as a color(vertical)-flavor(horizontal) mixed $2 \times 2$ matrix  and we have defined the following combinations:
\begin{align}
 &F_{\mu\nu} = F_{\mu\nu}^{a}\frac{\tau^{a}}2+F_{\mu\nu}^{0}\frac{\mathbbm 1_{2}}2, \quad A_{\mu}=A^{a}_{\mu}\frac{\tau^{a}}2, \quad \Tr (\tau^{a} \tau^{b})=\frac12 \delta^{ab}   \nonumber \\
  & \nabla_{\mu}Q= \partial_{\mu}Q-i g_{0}  A_{\mu}^{0}\frac{\mathbbm 1_{2}}2  Q-  i g A_{\mu} Q\,.
\end{align}
The potential in the model above is chosen as the most general quartic potential invariant under a $SU(2)$ flavor symmetry. The coefficients of the potential have to be chosen such that the model develops a non-trivial expectation value for the scalar fields, such that the gauge symmetry is completely Higgsed. This is achieved by choosing  $\lambda_{1}$, $\lambda_{2}$ and $m$ to be real positive. The last constant term is chosen  for convenience such that the energy density of the vacuum vanishes. The vacuum of the theory is color-flavor locked, and it is given by the following:
\beq
Q_{vev}=
\left(
\begin{array}{cc}
 \sqrt\xi & 0    \\
  0 &   \sqrt\xi
\end{array}
\right)\,,\quad \xi=\frac{m^{2}}{2(2\lambda_{1}+\lambda_{2})}\,.
\label{eq:vac}
\eeq
Together with the spontaneous breaking of the gauge symmetry, the crucial ingredient for the existence of non-Abelian vortices is the existence of a preserved diagonal color-flavor global symmetry in the vacuum~(\ref{eq:vac})
\begin{equation}
Q \rightarrow U_{C} Q U_{F}, \quad U_{C}=U_{F}^{\dagger}\,.
\label{eq:colflav}
\end{equation}
The symmetry breaking pattern of the model is then:
\begin{equation}
SU(2)_{C}\times  U(1)\times SU(2)_{F} \stackrel{\sqrt \xi}{\longrightarrow} SU(2)_{C+F}\,.
\end{equation}

 A minimal vortex configuration can be constructed as follows:
 \begin{align}
Q(r,\varphi)   &= 
\begin{pmatrix}
\phi_1(r) e^{i \varphi} & 0 \\
0 & \phi_2(r)
\end{pmatrix}  \nonumber \\
A_i+A_{i}^{0}\frac{\mathbbm 1}2 &= -\frac{1}{2}\epsilon_{ij}\frac{x^j}{r^2} \left[
f(r)\, \mathbf{1}_{2} + f_{\rm NA}(r) \, \tau^{3} \right] \ ,
\qquad  i=1,2 \ .  \label{eq:vortex}
\end{align}
This individual vortex solution violates the color-flavor locked symmetry and thus  develops orientational moduli and a corresponding moduli space:
\begin{equation}
\mathcal M=SU(2)_{C+F}/U(1)_{C+F}= \mathbb CP^{1}\,.
\label{eq:vortexsymm}
\end{equation}
 The low-energy excitations of these vortices  are thus described  by a $\mathbb CP^{1}$ NL$\sigma$M (non-linear sigma model). Non-Abelian vortices thus generalize the well-known Abrikosov-Nielsen-Olesen vortex solution~\cite{Abrikosov:1956sx,Nielsen:1973cs} with the inclusion of non-Abelian orientational degrees of freedom. 
 A model where the gauge symmetry is $SU(N)$,  $SO(2N)$ or $USp(2N)$ can be constructed similarly and their mathematical and physical properties 
 have been extensively studied in the last several years~\cite{Eto:2009bg,Ferretti:2007rp,Eto:2008qw,Eto:2011cv,Gudnason:2010jq,Gudnason:2010rm}. The detailed analysis of the case with arbitrary gauge groups~\cite{Eto:2008yi} is also possible in principle.

In the literature non-Abelian vortices have been often investigated  in the context of theories with extended $\mathcal N=2$ supersymmetry. The model of Eq.~(\ref{eq:act}) could be regarded as a bosonic sector of a theory with extended supersymmetry,  when a special choice is made for  the coefficients $\lambda_{1}$, $\lambda_{2}$ and $m$.  We will elaborate more on this in Section~\ref{secmoduli}.

\section{Weakly Gauged non-Abelian Vortices \label{subwgauge}}

\subsection{Weakly gauging flavor symmetry}
The global symmetry of the model~(\ref{eq:act}) is an $SU(2)_{F}$  flavor symmetry acting on the flavor index $A$  of the scalars $q^{A}$, or equivalently acting from the right on $Q$.  The main purpose of this paper is to study the effects on the non-Abelian vortex solutions, and their effective description, when we fully gauge this flavor symmetry. We rename it $SU(2)_R$, for convenience, while the original $SU(2)_{C}$ gauge symmetry acting on the left becomes $SU(2)_{L}$.  We assume that the fields $Q$ couple minimally to the newly introduced gauge fields, and we take the new gauging to be very weak:
\beq 
  g_{R} \ll   g_{L} = g.  
\label{eq:weak}
\eeq 
For simplicity we take also $g_{L} = g_{0}=g$, an equal coupling for $SU_{L}(2)$ and $U(1)$. This choice does not reduce the generality of the following discussions. 

In the following we discuss the case of $U(2)$ model but the generalization to the case of $U(N)$, $SO(2N)\times U(1)$ or $USp(2N)\times U(1)$ gauge theories is straightforward. The action is now 
\begin{eqnarray}
\mathcal L   & = &    \Tr\left[   \frac1{2 }(F_{\mu\nu}^{L})^{2}+\frac1{2 }(F_{\mu\nu}^{R})^{2}+ |\mathcal D_{\mu} Q|^{2} +\lambda_{2}(\bar Q  Q)^{2} -m^{2} \bar Q  Q  \right]+\lambda_{1}(\Tr[\bar Q Q])^{2}+\frac{m^{4}}{2(2\lambda_{1}+\lambda_{2})}\,,\nonumber \\
  \label{eq:gaugedact}
\end{eqnarray}
where 
\begin{align}
&F_{\mu\nu}^{L}=F_{\mu\nu}^{La}\frac{\tau^{a}}2+F_{\mu\nu}^{0}\frac{\mathbbm 1_{2}}2, , \quad F_{\mu\nu}^{R}=F_{\mu\nu}^{R\, c}\frac{\tau^{c}}2, \quad A_{\mu}^{L}=A^{La}_{\mu}\frac{\tau^{a}}2, \quad A_{\mu}^{R}=A_{\mu}^{R\, c}\frac{\tau^{c}}2& \nonumber \\
\nonumber \\
&  \mathcal D_{\mu}Q=\partial_{\mu}Q-i g  A_{\mu}^{0} \frac{\mathbbm 1_{2}}{2}  Q-  i g A_{\mu} Q-i g_{R} Q A^{R}_{\mu}\,.&
\end{align}
and the summation over the repeated $SU(2)$ generator indices $a,c=1,2,3$ is implicit as usual. We use the same potential as in the previous Section, since this is already the most general compatible with the symmetries of the theory. Moreover, the vacuum of the theory is still given by the equation~(\ref{eq:vac}). 
The critical new element we have to consider when fully gauging the flavor symmetry is the presence of an unbroken gauge symmetry. The unbroken, global color-flavor symmetry discussed in the previous Section becomes now a gauge symmetry:
\beq
SU(2)_{L}\times U(1) \times SU(2)_{R}\rightarrow SU(2)_{V}\,.	
\eeq

The model we are considering  still possesses non-trivial vortex solution supported by the breaking of the $U(1)$ gauge group. However, vortices are now coupled to gauge fields  which are massless in the bulk. Stable vortex configurations can be constructed for generic values of the gauge couplings $g_{L}$ and $g_{R}$. However, we exploit  the condition~(\ref{eq:weak}) to study the effects of a weak coupling on the non-Abelian vortex physics. Let us then change variables 
\begin{eqnarray}   
&A_{\mu}^{L} =  c\,  B_{\mu} -  s\,  V_{\mu},\quad    A_{\mu}^{R} =   c\,  V_{\mu} +  s\, B_{\mu}, & \nonumber \\
  \nonumber \\
  &c\equiv \cos \alpha,  \quad s\equiv \sin \alpha, &
\end{eqnarray}
where $\alpha$ is a mixing angle to be determined shortly. Inverting the expressions above we get
\beq   V_{\mu}=  c\,  A_{\mu}^{R} -   s\,  A_{\mu}^{L},  \quad  B_{\mu} =  c\,  A_{\mu}^{L} +  s\,  A_{\mu}^{R}\;.
\eeq
where $ V_{\mu} $ and $ B_{\mu} $ stand for  the ``vector-like'' and ``broken''  gauge fields (below).  We determine the mixing of  $A_{\mu}^{L}$ and $A_{\mu}^{R}$   so that the vacuum 
\beq       
Q_{vev} =  \sqrt\xi\, {\mathbbm 1_{2}} 
\eeq
respects $SU(2)_{V}$, i.e., leave $ V_{\mu}$ massless. From the covariant derivative we extract the mass terms for the gauge fields:
\beq   
g  A^{L}_{\mu} Q_{vev} + g_{R} Q_{vev}  A_{\mu}^{R} \propto g A_{\mu}^{L} + g_{R}   A_{\mu}^{R}=  g \, (c\,  B_{\mu}  -   s\,  V_{\mu})   +  g_{R}\,( c\,  V_{\mu} +  s\,  B_{\mu}) \,.       
\eeq
The vanishing of the term proportional to $V_{\mu}$ implies the following relations
\beq   c\, g_{R} = s\, g , \qquad    \cos \alpha =   \frac{g}{\sqrt{g^{2}+ g_{R}^{2}}}, \quad   \sin \alpha =   \frac{g_R}{\sqrt{g ^{2}+ g_{R}^{2}}} \,.
\eeq
In the weak coupling limit  $\alpha \ll 1$, 
\beq    c=\cos \alpha \simeq 1, \quad s=\sin \alpha \ll 1\;. \eeq
In the following we will  expand all our formulas keeping only the linear terms in $\alpha$. Now, if we define
\beq      g_{V} \equiv c\, g_{R} = s\, g=   \frac{g g_{R}}{\sqrt{g^{2}+ g_{R}^{2}}}   \simeq  g_{R}, 
\eeq
 up to linear terms in $\alpha$ (equivalently, the ratio $g_{R}/g_{L}$) we have, for the term in the covariant derivative:
\beq   g  A^{L}_{\mu} Q + g_{R} Q  A_{\mu}^{R} \simeq    g \,  B_{\mu} \, Q  -  g_{V} \, [  V_{\mu}, Q]\,, 
\eeq
which makes manifest the fact that the field $V_{\mu}$ is massless on the vacuum~(\ref{eq:vac})\footnote{The fields $V_{\mu}$ and $B_{\mu}$ are defined here for convenience only. They cannot be consistently defined at the same time for the bulk theory as gauge fields.   However, this is legitimate once we are interested on the effective theory on the vortex, where most of the gauge symmetry is broken. We will be left with the gauge fields $V_{\mu}$ and the diagonal gauge symmetry only, as a (gauge) symmetry for the worldsheet theory of the vortex.}. In fact, defining
\beq     B_{\mu}  = \left(\begin{array}{cc}B_{\mu}^{ 3} & B_{\mu}^{\, -} \\ B_{\mu}^{+} &-B_{\mu}^{3}\end{array}\right) 
\eeq
we have the following mass terms:
\beq 
\Tr \,  |g A_{\mu}^{{0} } + g B_{\mu} Q -  g_{V} \, [  V_{\mu}, Q]|^{2}= 2 g^{2}\xi  \,[\, ( A_{\mu}^{0})^{2}+  (B_{\mu}^{3})^{2} +   B_{\mu}^{-} B_{\mu}^{+}\,]\,.
\eeq
 $A^{(0)}$ and  $B^{i}$\,  ($i=1,2,3$) are all massive with mass $g\, \sqrt\xi$.

So far, we have discussed the mass terms of the gauge fields in the vacuum of the theory. Let us discuss what happens now in the presence of a vortex. As we have already recalled in the previous section, a vortex solution would break the $SU(2)_{C+F}$ color-flavor symmetry of the vacuum in the model without gauging of the flavor symmetry to a residual $U(1)$. In the weakly gauged model we consider in this paper, the color flavor symmetry is now gauged to a diagonal gauge symmetry, which is broken to a $U(1)$ residual symmetry on the vortex. The main effect of putting a vortex in our weakly gauged model is then to induce a spatially dependent effective mass term confined on the vortex solution.  A  vortex solution in a fixed orientation  has the  $Q$ fields as in Eq.~(\ref{eq:vortex}). We can thus evaluate the mass terms on a vortex solution as follows (suppressing the contracted  Lorentz indices for the vector fields):
 \begin{align}  &\frac1{\xi}\Tr \, | g\,A^{0 } Q+  g \,  B \, Q  -  g \, [  V, Q]  |^{2} = \nonumber \\
 & =\Tr \left| \left(\begin{array}{cc}g (A^{0}+B^{3})\phi_{1} & g  B^{-}\phi_{2} - g_{V} \, V^{ -}(\phi_{2}-\phi_{1}) \\ g  B^{+}\phi_{1} + g_{V} \, V^{ +}(\phi_{2}-\phi_{1})  & g (A^{0} -B^{3})\phi_{2} \end{array}\right) \right|^{2}=\nonumber \\
 &=  g^{2}  (A^{0} +B^{3})^{2}\phi_{1}^{2}+ g^{2}  (A^{(0)} -B^{3})^{2}\phi_{2}^{2}  +  g_{V}^{2} V^{-} V^{+} (\phi_{1}-\phi_{2})^{2}+\nonumber \\
 & -g g_{V}(B^{-}V^{+}+V^{-}B^{+})(\phi_{1}-\phi_{2})^{2}+ g^{2}B^{-}B^{+}(\phi_{1}^{2}+\phi_{2}^{2})\,.\nonumber \\
 \end{align}
 From the expression above we see that all the gauge fields acquire mass on a vortex solution where $\phi_{1}\neq\phi_{2}$, except the field $V^{3}$ which remains massless as already anticipated. At the vortex core,
 \beq   Q_{core}=   \left(\begin{array}{cc}0 & 0 \\0 & \phi_{2}(0)\end{array}\right), \quad \phi_{2}(0)\neq0\;, 
\eeq
and the mass terms become proportional to:
\begin{align}
& g^{2}  (A^{0} -B^{3})^{2}   +   \left(g^{2}  B^{- }B^{+} -  g g_{V}\, (B^{-} V^{+}+B^{+} V^{-}) +   g_{V}^{2} V^{-} V^{+}\right).  \label{SU(2)}
 \end{align}
Diagonalization of the last term shows that there is a small mixing between $ B^{\pm}$ and $ V^{\pm}$ 
 \beq    B^{core\, \pm}\sim  B^{\pm} -  \tfrac{g_{V}}{g} V^{\pm}; \qquad V^{core \, \pm}\sim V^{\pm} +  \tfrac{g_{V}}{g} B^{i}
 \eeq
 with the corresponding masses as follows 
 \beq    m_{B^{\pm}}^{core} \simeq   g\, \sqrt\xi; \qquad m_{V^{\pm}}^{core} \simeq   g_{V} \sqrt\xi\;
 \eeq
 that is, $V^{\pm}$ gets a small mass of order $g_{V}$.
 
  Of the two massive fields in the bulk  $A^{0}$, $B^{3}$, the combination  $A^{0} - B^{3}$ remains massive, while the other combination  $A^{0} +  B^{3}$
  becomes massless. The vanishing of the effective mass term for a $U(1)$ gauge field is a well-known effect for both Abelian and non-Abelian vortices, and is ultimately the effect thanks to which superconductivity is broken at the vortex core,  allowing the  magnetic flux to penetrate near the core. The effect we study in our paper has thus the opposite effect for the newly introduced gauge symmetry.

  Recapitulating, apart from the small mixing near the vortex core,  the fields $A^{0}$ and $B$ are the broken (Higgsed) gauge fields 
  which make up the vortex configurations, together with the scalar fields.  The combination  $A^{0} +B^{3}$ gets restored along the vortex core: it gives rise to the nonzero magnetic flux
  carried by the vortex.  The vector-like  field $V$ are  the gauge fields which are  ``unbroken'' in the bulk.  On the vortex core, the non-diagonal components $V^{\pm}$,  acquire small masses   (Higgsed slightly), whereas the component  $V^{3}$ remains unbroken, both in the bulk and in the vortex region. 
  
  What happens is that the components $V^{\pm}$, absorb the orientational zero-modes of the vortex and become massive. However as their mass $\mathcal O( g_{V} \sqrt\xi )$ is much smaller than 
  the typical  mass scales  $\sqrt\xi$, $g \sqrt\xi$ of the vortex, they  describe recognizable,  light physical excitations along the non-Abelian vortex. 
It makes perfect sense to ask how such excitations are described in a modified effective action.
    
\subsection{Effective action}

The static vortex configurations will be only slightly modified by the introduction of the right gauge fields, when the latter coupling is very small.
 Our main interest below will be to show how the vortex effective action (sigma model) describing the  fluctuation of the $\mathbb C P^{1}$ orientational zero-modes, gets modified.  We adopt the formalism introduced in Ref.~\cite{Gudnason:2010rm}, which is the most effective for a general discussion. The final results apply equally well with a trivial generalization  to the cases of vortices in $SU(N)$, $USp(2N)$ and $SO(N)$ gauge theories. 

As usual, the starting point is a static vortex configuration chosen in some arbitrary orientation obtained through a generic rotation $U$ of the solution~(\ref{eq:vortex}).
In the singular gauge we have:
\begin{align}
& Q   = U
\begin{pmatrix}
\phi_1(r) & 0 \\
0 & \phi_2(r)
\end{pmatrix} U^{-1}
= \frac{\phi_1(r)+\phi_2(r)}{2}\,  \mathbf{1}_{2}
+ \frac{\phi_1(r)-\phi_2(r)}{2} \,U T U^{-1} \ , \non
&A_{i}^{0}\frac{\mathbbm 1_{2}}{2}+B_i = -\frac{1}{2}\epsilon_{ij}\frac{x^j}{r^2} \left[
f(r)\, \mathbbm{1}_{2} + f_{\rm NA}(r) \, U T U^{-1} \right] \ ,
\qquad  i=1,2 \ ,\nonumber \\
\nonumber \\
& T  = {\rm diag}\,\left(1, -1\right) = \tau^{3} \,.
  \label{genericorient}
\end{align}
In the expression above we have already taken into account the weak gauging of the right symmetry by constructing the vortex in terms of the mixed gauge field $B_{i}$. 
In the following, we will use the convenient explicit parameterization of $U$: 
\begin{align}
U =
\begin{pmatrix}
\mathbf{1} & - B^\dag \\
0 & \mathbf{1}
\end{pmatrix}
\begin{pmatrix}
X^{-\frac{1}{2}} & 0 \\
0 & Y^{-\frac{1}{2}}
\end{pmatrix}
\begin{pmatrix}
\mathbf{1} & 0 \\
B & \mathbf{1}
\end{pmatrix}
=
\begin{pmatrix}
X^{-\frac{1}{2}} & - B^\dag Y^{-\frac{1}{2}} \\
B X^{-\frac{1}{2}} & Y^{-\frac{1}{2}}
\end{pmatrix}
\label{eq:Umatrix} \ ,
\end{align}
with  $X$ and $Y$ defined by
\beq
X\equiv\mathbf{1} + B^\dag B \ , \quad
Y\equiv\mathbf{1}  + B B^\dag   \label{X&Y}\,.
\eeq
In the case of the $SU(2)$ theory, $B$ is just a complex number which corresponds the inhomogeneous coordinate of $\mathbb C P^{1}$.

  Before proceeding let us briefly mention the modifications needed in the cases of more general gauge theories.  In the case of the $SU(N)$ theory $B$ becomes a column vector
\beq B =
\begin{pmatrix}
b_1 \\ \vdots \\ b_{N-1}
\end{pmatrix} \ , \label{eq:B_CPNvector} \eeq
while $B^{\dagger}$ is correspondingly a row-vector. Moreover
\beq X = 1 + B^\dag B \ , \qquad Y = \mathbf{1}_{N-1} + B B^\dag \ ,
\label{eq:XY_CPN} \eeq
are a scalar and an $(N-1)\times (N-1)$ dimensional matrix,
respectively.
$B$ represents the standard inhomogeneous
coordinates of $\mathbb{C}P^{N-1}$, and the matrix $T$ in~(\ref{genericorient})  take the form, 
$  T = diag (1, - {\mathbbm 1}_{N-1})$.  The unit matrices in $U$ take appropriate dimensions.  
The relation between the $N$-component complex unit
vector $n$ used by Gorsky \emph{et. al.} in Ref.~\cite{Gorsky:2004ad} and our $B$
matrix is given by
\beq n =
\begin{pmatrix}
X^{-\frac{1}{2}} \\
B X^{-\frac{1}{2}}
\end{pmatrix} \ , \label{identity} \eeq
and is discussed in more detail in Appendix~\ref{sec:Bnconn}.
In the case of $USp(2N)$ and $SO(2N)$  theories,  the coordinates $B$ are 
symmetric or antisymmetric complex $N\times N$ matrices, respectively, and the unit matrices will  have all $N\times N$ dimensions.

The standard procedure to obtain an effective action for the low energy physics is to allow the moduli parameters $B$ to fluctuate on the vortex worldsheet: 
\beq B = B(x^\alpha) \ , \qquad  x^\alpha =(x^3,x^0) \ . \eeq
Evaluating the action~(\ref{eq:act}) on the vortex configuration with the $x_{\alpha}$ dependence above, we obtain a static (tension) term plus the following relevant term for the world-sheet coordinates: 
\beq
 \Tr \sum_{\alpha=0,3}
\left[  |\p_\alpha Q|^{2} +
\sum_{i=1,2}  |F_{i \alpha}|^{2}  \right] \ ,
\label{eq:large1}
\eeq
however, this term suffers from a divergence coming from the vortex core.  When there is no right gauge field coupling, $g_{R}=0$, we know how to treat this problem. In order to extract the minimum-energy excitation, one must 
introduce the longitudinal gauge field components
\begin{align}
A_\alpha = i\, \rho(r)\,
U \left( U^{-1}\p_\alpha U \right)_{\perp} U^{-1} \ , \qquad
\alpha=0,3\ , \nonumber \\
\left( U^{-1} \partial_\alpha U \right)_{\perp} \equiv \frac{1}{2}  \, [\,  U^{-1} \partial_\alpha U - T\,  U^{-1} \partial_\alpha U \, T \,]\;.  
\label{eq:AalphaAnsatz}
\end{align}
and compute instead 
\beq
\Tr \sum_{\alpha=0,3}
\left[  |(\de_\alpha  - i g A_{ \alpha} )  Q|^{2} +
\sum_{i=1,2}  |F_{i \alpha}|^{2}  \right] \ .\qquad  
\label{eq:largecured}
\eeq
One has then to minimize the above expression in order to determine the vortex profiles $\phi_{1}, \phi_{2},  f, f_{NA},\rho$.  

In the presence of a weakly coupled vector gauge fields $V_{\mu}$, unbroken in the bulk and very slightly broken in the vortex core,  we generalize the above procedure as follows. We have to compute the following term:
\beq
\mathcal L_{eff}= \Tr \sum_{\alpha=0,3}
\left[ | \nabla_{\alpha} Q -  i g  B_{\alpha} Q |^{2} +
\sum_{i=1,2}  |F^{L}_{i \alpha}|^{2}+\sum_{i=1,2}  |F^{R}_{i \alpha}|^{2}  \right] \ .\qquad  
\label{eq:gaugedaction}
\eeq
The scalar kinetic terms   are replaced by:
\beq     
\Tr |(\de_\alpha  - i g A_{L\, \alpha} ) Q |^{2} \longrightarrow  \Tr\, | \nabla_{\alpha} Q -  i g  B_{\alpha} Q |^{2}, 
\eeq
where we have defined:
\beq
 &\nabla_{\alpha} Q = (\de_{\alpha} +  i g_{V} [V_{\alpha}, \cdot  ) Q  \nonumber \\
   &B_\alpha = i\, \rho(r)\,
U \left( U^{-1} \nabla_\alpha U \right)_{\perp} U^{-1} \ , \quad  \nabla_{\alpha} U  = (\de_{\alpha} +  i g   V_{\alpha})   U\,,\nonumber \\
 \nonumber \\
  & \left( U^{-1} \nabla_\alpha U \right)_{\perp} \equiv \frac{1}{2}  \, [\,  U^{-1} \nabla_\alpha U - T\,  U^{-1} \nabla_\alpha U \, T \,]\;.  
\label{AnsatzGene}
\eeq

Let us now consider now the  gauge field tensor terms,  
\begin{align}
    F_{\mu\nu}^{L} &=  \de_{\mu}  (c\,  B_{\nu} -  s\,  V_{\nu} ) -    \de_{\nu}  (c\,  B_{\mu} -  s\,  V_{\mu})  - i g  [c\,  B_{\mu} -  s\,  V_{\mu}, c\,  B_{\nu} -  s\,  V_{\nu} ] = \nonumber \\
& =     \{ \de_{\mu}  + i  g_{V}  [V_{\mu} \cdot  \} B_{\nu}  -   \{ \de_{\nu}  + i  g_{V}  [V_{\nu} \cdot  \} B_{\mu}  - i g   [ B_{\mu},  B_{\nu}] -\frac{g_{V}}{g}F^{V}_{\mu\nu} \equiv \nonumber \\
&\equiv  F_{\mu\nu}^{B_{\nabla_{+}}}-\frac{g_{V}}{g}F^{V}_{\mu\nu}\,;\nonumber \\
\nonumber \\
   F_{\mu\nu}^{R} &=  \de_{\mu}  (c\,  V_{\nu} +  s\,  B_{\nu} ) -    \de_{\nu}  (c\,  V_{\mu} + s\,  B_{\mu})  - i g  [c\,  V_{\mu} +  s\,  B_{\mu}, c\,  V_{\nu}+  s\,  B_{\nu} ]= \nonumber \\
& =    F_{\mu\nu}^{V} +  \frac{g_{V}}{g}\left\{ \left(\de_{\mu}  - i  g_{V}  [V_{\mu} \cdot  \right)B_{\nu}  -   \left( \de_{\nu}  - i  g_{V}  [V_{\nu} \cdot  \right) B_{\mu}  - i g   [ B_{\mu},  B_{\nu}]  \right\} \equiv \nonumber \\
&\equiv  F^{V}_{\mu\nu}+\frac{g_{V}}{g} F_{\mu\nu}^{B_{\nabla_{-}}}\,.
 \label{covariant} 
\end{align}
The sum of the corresponding kinetic terms becomes, when we keep only terms up to the first order in $g_{V}$:
\begin{equation}
  (F^{L}_{\mu\nu})^{2}+ (F^{R}_{\mu\nu})^{2} =( F_{\mu\nu}^{B_{\nabla_{+}}})^{2}+( F_{\mu\nu}^V)^{2} 
\end{equation}

The final form of the action~(\ref{eq:gaugedaction}) reduces then to the following
\beq
\mathcal L_{eff}= \Tr
\left[ | \nabla_{\alpha} Q -  i g  B_{\alpha} Q |^{2} +
  (F^{B_{\nabla_{+}}}_{i \alpha})^{2}+  (F^{V}_{\mu\nu})^{2}  \right] \ ,  
\label{eq:gaugedactionfinal}
\eeq
where all the indices are summed. Notice that in the first two terms we keep only terms up to two derivatives of the worldsheet coordinates $\alpha$, as usual in this type of calculations. The last term however describes gauge fields not confined on the vortex, and then we keep explicit all the indices $\mu,\nu$. This is analogous to the case studied by Witten for the Abelian superconducting string~\cite{Witten:1984eb}.  The calculation of the first two terms now proceed straightforwardly, although lengthily. We closely follow  Ref.~\cite{Gudnason:2010rm}. First we have to  check the following orthogonality conditions
\beq  
& \Tr\, (U^{-1} \nabla_{\alpha} U)_{\perp}  \, q|^{U=0})=0, \qquad  \qquad  \Tr\,[\, U (U^{-1} \nabla_{\alpha} U)_{\perp}U^{-1}  \, q \,] =0&,\nonumber \\
\nonumber \\
  &  \Tr\, [\,U (U^{-1} \nabla_{\alpha} U)_{\perp}U^{-1}  \, \nabla_{\alpha}  q \,]=0\;. &
\eeq
Using the conditions above we  eventually find the following expressions
 \begin{align}
\Tr\, |{\cal D}_\alpha \, Q|^2 &= -  \left[
\frac{\rho^2}{2}\left(\phi_1^2+\phi_2^2\right)
+(1-\rho)\left(\phi_1-\phi_2\right)^2\right]
\Tr \left[\left( U^{-1}\nabla_\alpha U \right)_{\perp}\right]^2 \ , \nonumber \\
\Tr(F^{B_{\nabla_{+}}}_{i \alpha})^{2}&= - \left[
\left(\p_r\rho\right)^2
+\frac{1}{r^2}f_{\rm NA}^2\left(1-\rho\right)^2\right]
\Tr \left[\left( U^{-1}\nabla_\alpha U \right)_{\perp} \right]^2 \ .
\end{align}
Notice that the result above is similar to the one that one would obtain in the case without right gauging. The only difference is the covariant derivative instead of a normal derivative appearing on the left. This result has to be expected, because of the vector gauge invariance unbroken in the vacuum. We stress in fact that only the terms up to the linear order in $g_{V}$ have been consistently carried out through the entire calculation. However, because of gauge invariance we can reintroduce the quadratic terms appearing in the covariant derivatives above.

After integrating over the transverse plane $x,y$ and minimizing with respect to  the profile functions  $\phi_{1}, \phi_{2},  f, f_{NA}, \rho$,  
we obtain the main result of this paper, the effective action for a weakly gauge non-Abelian vortex:  
\begin{align}
\nonumber \\
S  
=  2\beta \int dtdz \;& \Tr\left\{
\left(\mathbbm 1 + B^\dag B\right)^{-1} (\hat{\nabla}_\alpha B)^\dag
\left(\mathbbm{1} + B B^\dag\right)^{-1}\hat{\nabla}_\alpha B
\right\} - \frac{1}{2} \int  d^{4}x\,  \Tr (F^{V\, }_{\mu \nu})^{2}=   \nonumber \\
\nonumber \\
&= 2\beta \int dtdz \; \Tr
 \frac{|\hat{\nabla}_\alpha B|^{2}}{\left(1 + |B|^{2}\right)^{2}}
 - \frac{1}{2} \int  d^{4}x\,  \Tr (F^{V\, }_{\mu \nu})^{2} \,.
\nonumber \\
 \label{eq:efffinal}
\end{align}
We have written the result in a matrix-like form appropriate when generalizing to higher rank gauge groups, while in the second line we specialized  it to the $SU(2)$ case.  $\hat{\nabla}_\alpha B$ is a covariant derivative defined in Eq.~(\ref{defined1}) in the general case. In the SU(2) case, it becomes:
\beq    \hat{\nabla}_{\alpha} B  \equiv \de_{\alpha}  B  - i g_{V} ( 2 V_{\alpha}^{3} B   - V_{\alpha}^{+} +    V_{\alpha}^{-} B^{2} )\;,
\label{defined11}\eeq
This form of the covariant derivative reflects the fact that $B$ transforms non-linearly under the unbroken $SU(2)$ diagonal group. In the expression above we have obtained as effective action a gauged non-linear sigma model in two dimensions coupled to non-Abelian gauge fields propagating in  four-dimensions.   
 The coupling is given  by
\begin{align}
\beta = \frac{2\pi}{g^2} \mathcal I.    \label{beta}
\end{align}
with the constant $\mathcal I$ given by the usual integral~\cite{Auzzi:2003fs,Shifman:2004dr}
\begin{equation}
\mathcal I= \int dr \, r \left[
\frac{\rho^2}{2}\left(\phi_1^2+\phi_2^2\right)
+(1-\rho)\left(\phi_1-\phi_2\right)^2+\left(\p_r\rho\right)^2
+\frac{1}{r^2}f_{\rm NA}^2\left(1-\rho\right)^2\right]\,,
\end{equation}
that has to be evaluated numerically in the general case, while in  the BPS case  (discussed in Section~\ref{secmoduli} below)   it reduces to a total derivative and can be evaluated exactly:
\beq
\mathcal I_{BPS}=1\,.
\eeq
We see from Eq.~(\ref{defined11}) that an appropriate combination of $V_{\alpha}$ absorbs the zero-modes $\de_{\alpha}  B$  and gets mass $g\, \sqrt\xi$.  Near $B=0$ it is simply the non-diagonal 
gauge bosons  $ V_{\alpha}^{\pm}$ that get mass, consistently with Eq.~(\ref{SU(2)}).

In the case of $SU(N)$ theory the first term corresponding to  the ``gauged $\mathbb CP^{N-1}$ action''
can be rewritten in  terms of the homogeneous vector field $n$ of unitary length 
\beq n =
\begin{pmatrix}
X^{-\frac{1}{2}} \\
B X^{-\frac{1}{2}}
\end{pmatrix} \ ,   \label{identi3} \eeq
 which transforms linearly under the unbroken gauge symmetry:
 \begin{align}
\nonumber \\
&S  = 2\beta \int dtdz \left( \nabla_{\alpha}n^{*}\cdot\nabla_{\alpha}n+(\nabla_{\alpha}n^{*}\cdot n)^{2}\right)
 - \frac{1}{2} \int  d^{4}x\,  \Tr (F^{V\, }_{\mu \nu})^{2} \,,\nonumber \\
 \nonumber \\
 & \nabla_{\alpha} n  = (\de_{\alpha} +  i g_{V}   V_{\alpha})   n\,.
 \label{eq:lineareff}
\end{align}

 Notice that the number of orientational zero-modes  $B$  in general cases, $2(N-1)$, $N(N-1)$ and $N(N+1)$, respectively, for   $\mathbb CP^{N-1}$,   Hermitian symmetric spaces, $SO(2N)/U(N)$ and $USp(2N)/U(N)$,    correspond exactly to the number of gauge fields of the broken symmetries.  We thus see that all the orientational degrees of freedom are eaten by the weak Higgs mechanism on  the vortex.

\section{Decay of the Massive $V^{\pm}$ Bosons}

\begin{figure}
\begin{center}
\includegraphics[width=4in]{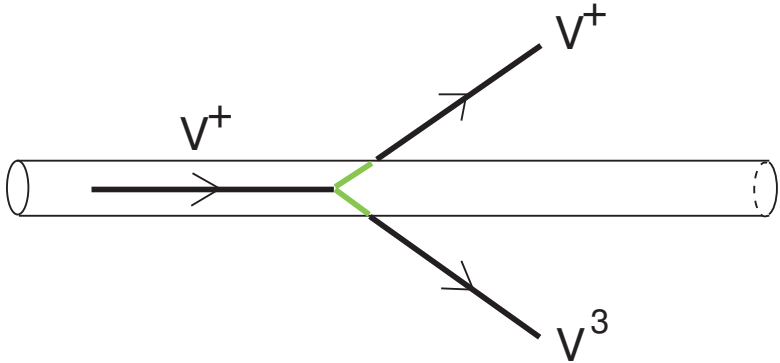}
\caption{Schematic representation of the decay process of a gauge boson mediated by a non-Abelian vortex.}
\label{DecayV}
\end{center}
\end{figure}

 The presence of a Higgs mechanism, and thus mass generation, localized on a topological soliton has a novel consequence for the light, but massive, particles propagating along the vortex:  the $V_{\mu}$ particles cannot be stable. The point is that they decay through a process of the type 
\beq     V^{\pm}  \to   V^{\pm} + V^{3}\;,\label{process}
\eeq
via the cubic non-Abelian gauge couplings in Eq.~(\ref{eq:efffinal}).   If the process is to occur entirely within the vortex core,   the $V^{\pm}$ remains  massive and it is kinematically forbidden.    On the other hand, 
both  $V^{\pm}$ and  $V^{3}$ are massless outside the vortex and the process is perfectly allowed if the final-state particles are in the bulk (Fig.~\ref{DecayV}).  In other words,~(\ref{process}) is a tunneling  process. 
The tunneling probability can be roughly estimated assuming the energy and momentum are conserved when the particles escape from the vortex.  The energy barrier is then about the mass of the particles in the vortex, since we have to account for creating a particle and some momentum both of order of the mass 
\beq    \Delta E \sim 2 M_{V}  = 2 g_{V}\, \sqrt\xi.
\eeq
The distance the virtual particles $V^{\pm,3}$  must travel under the barrier is given by thickness of the vortex
\beq    L \sim \frac{1}{g \, \sqrt\xi}.
\eeq
The tunneling probability is then
\beq    P \sim e^{-\Delta E L}\sim  e^{ - 2 g_{V} /g} \sim 1\,, \qquad   g \ll  g_{L}\;,
\eeq
and the virtual particles can easily escape from the vortex.

The process~(\ref{process}) is thus suppressed only by the small coupling constant $g_{V}$ (small decay amplitude) and by a small phase space.  From a dimensional consideration 
the decay probability is of the order of 
\beq   \Gamma \sim   g_{V}^{2} \, M_{V}  =   g_{V}^{3} \, \sqrt\xi\;.  
\eeq

It is somewhat curious that the physical excitation of orientational modes in the vortex worldsheet gets diffused in the bulk and disappears, while 
without the gauging of the flavor symmetry, the former represents  stable, massless modes (Nambu-Goldstone particles)  propagating in the worldsheet.  The massless particles are confined in the vortex core as they become massive in the bulk. There is thus an amusing inversion of the  situation: before the $SU(2)_R$ gauging the $B$ field is massless and confined in the vortex;  it cannot go out of the vortex into the bulk where it becomes massive.  With the weak gauging,  the $B$ fields are first eaten by the $V$ fields and become  massive inside the vortex, and then 
tend to escape into the bulk where they become massless, a sort of inverse Meissner effect.

 \section{BPS Limit and Supersymmetry \label{secmoduli} }
  
  \subsection{Moduli matrix}
  
 In this Section we consider more specifically the BPS saturated case. This happens with the following special choice of parameters:
\begin{equation}
\lambda_{2}=\frac{g_{L}^{2}+g_{R}^{2}}{4},\quad m^{2}=\frac{g_{L}^{2}}{2}\xi,\quad  \lambda_{1}=-\frac{g_{R}^{2}}{8} 
\end{equation}
 the potential of Eq.~(\ref{eq:act}) then reduces to the following:
\begin{equation}
V_{SUSY}=\Tr\left\{\frac{g_{L}^{2}}{4}\left(\bar Q Q-\xi \right)^{2 }+\frac{g_{R}^{2}}{4}\left(\bar Q Q-\frac{\mathbbm 1}{2} \Tr \bar Q Q \right)^{2 } \right\}\,.
\end{equation}
Such a BPS vortex naturally arises in the bosonic sector of ${\cal N}=2$  supersymmetric SQCD.   
The supersymmetric model involves also fields in the anti-fundamental representation $\tilde Q$ and an adjoint scalar, all in chiral multiplets containing fermion partners. These additional fields are however trivial on the background of a BPS vortex, and can be discarded when constructing semi-classical configurations. The additional fields are however crucial when one considers quantum effects.
 
This special choice dramatically simplifies the study of vortex solutions. First of all, the action can be rewritten by the Bogomol'nyi completion~\cite{Bogomol'nyi:1975de}:
\begin{eqnarray}
S & = & \int d^{2}x\, \Tr\left\{  \left[ F^{L}_{12}+ \frac{g_{L}}2( Q  \bar Q -\xi)\right]^{2} +  \left[ F^{R}_{12}+ \frac{g_{R}}2   \left( \bar Q Q -     \frac{1}{2}  \Tr (\bar Q Q) {\mathbbm 1}  \right)   \right]^{2}+\right. \nonumber \\
 & + &\left. |\nabla_{1} Q+i \nabla_{2} Q|^{2} + g_{L}  \, \xi\,F_{12}-\frac12\epsilon_{ij}\partial_{i}\left(i \nabla_{j}Q\bar Q- Q\,i\nabla_{j}\bar Q \right) \right\}.
 \label{eq:boghiggs}
\end{eqnarray}
As usual, this allows us to reduce the equations of motion to  first order differential equations:
\beq
&& F^{L}_{12}=- \frac{g_{L}}{2} ( Q \bar  Q -\xi) \label{BPS1}  	\nonumber \\
 &&F^{R}_{12} =  - \frac{g_{R}}2    \left( \bar Q Q -     \frac{1}{2}  \Tr (\bar Q Q) {\mathbbm 1}  \right) \label{BPS2}   \nonumber \\
 && \nabla_{1} Q+i \nabla_{2} Q  =  0\;.\label{BPSeq} 
\eeq
 Moreover, vortices are 1/2 BPS saturated, meaning that they preserve 1/2 half of the bulk supercharges. The presence of a residual supersymmetry, has allowed to derive exact quantitative results about the strongly-coupled physics of the vortex zero-modes~\cite{Hanany:2003hp,Hanany:2004ea,Shifman:2004dr,Dorey:1999zk,Dorey:1998yh}, and to put the in correspondence to those available for the bulk theory by using the exact Seiberg-Witten  solutions~\cite{Seiberg:1994rs,Seiberg:1994aj}.

For the static vortex configurations   the well-consolidated techniques of moduli matrix can be used to get further information~\cite{Eto:2006pg,Isozumi:2004vg,Eto:2005yh,Eto:2006uw,Eto:2006cx,Eto:2006db}.  As usual, we can solve  equations~(\ref{BPSeq}) by using a   (``moduli'') matrix  $H_{0}$ holomorphic in $z= x + i y$. The last of these is solved by 
\beq
Q(z,\bar z)=S^{-1}(z,\bar z)H_{0}(z)S_{R}^{-1}(z,\bar z)\;,  \nonumber \\
A_{\bar z}=\frac{ i}{g_{L}}  S^{-1} \partial_{\bar z} S, \qquad  A^{R}_{\bar z}= \frac{i}{g_{R}} (\partial_{\bar z} S_{R})   S_{R}^{-1} 
\label{bpssol}\eeq
whereas the first two  become 
\begin{eqnarray}
&&  4\, \partial_{\bar z}\left(\Omega\partial_{ z}\Omega^{-1}\right)+g_{L}\left(    H_{0}\,\Omega^{-1}_{R}H_{0}^{\dagger}\Omega^{-1} -\xi   \right)=0,\qquad \Omega \equiv S \, S^{\dagger }\;,    \\
&&  4\, \partial_{\bar z}\left(\Omega_{R}^{-1}   \partial_{ z}\Omega_{R}   \right)+g_{R}\left( \Omega_{R}^{-1} H_{0}^{\dagger}\Omega^{-1} H_{0}  -   \frac{{\mathbbm 1}  }{2} 
  \Tr   ( \Omega_{R}^{-1}  H_{0}^{\dagger}\Omega^{-1} H_{0} )  \right)=0,\qquad \Omega_{R} \equiv S_{R}^{\dagger } \, S_{R}\;, 
\nonumber   \end{eqnarray}
where the matrices  $S$ and $S_{R}$ are complexified  $U(2)_{L}$ and $SU(2)_{R}$ group elements;  they belong to $GL(2,\mathbb C)$ and $SL(2, \mathbb C)$, respectively. 
As usual, the separation in the holomorphic  moduli matrix and $S$ and $S_{R}$ matrices are defined only {\it modulo} a ``$V$-equivalence'' 
\beq   H_{0} \to   V(z)  \, H_{0}\,  V_{R}(z), \qquad   S \to V(z)  S, \qquad  S_{R}  \to  V _{R}\, S_{R}
\eeq
where holomorphic matrices $V$ and $V_{R}$ belonging to  $GL(2,\mathbb C)$ and $SL(2, \mathbb C)$, respectively, define the equivalence relations.

For a minimum winding vortex the condition on the moduli matrix is, as usual,
\beq
\det H_{0}=z-z_{0}    \,.
\eeq
When the $SU(2)_R$ symmetry is not gauged, the non-equivalent set of  solutions are labeled by the moduli matrix of the form
\beq
H_{0}=
\left(
\begin{array}{ccc}
  1& B   \\
  0& z-z_{0} 
\end{array}
\right)\,.
\eeq
where the complex number $B$ represents the $\mathbb C P^1$ coordinate, as well-known. 
The situation, in the presence of the right $SU(2)$ gauge fields coupled to the system, is entirely different.
 We have now a $V_{R}$ transformation acting on the right at our disposal, together with a left $V$
transformation.  It can be easily seen that $B$ can be chosen to have any value. In other words, there are no orientational moduli at all, 
only a position modulus remains. We can choose the  value 0 for convenience:
\beq
H_{0}=
\left(
\begin{array}{ccc}
  1& 0 \\
  0& z-z_{0} 
\end{array}
\right)\,.  \label{unique}
\eeq
Eqs.~(\ref{eq:boghiggs})-Eq.~(\ref{unique})  strongly suggest that there is indeed a unique vortex solution with tension 
\beq   T =   2\pi  \xi\,,   \label{tension}
\eeq
the same as that of the non-Abelian vortex solutions in the absence of the right gauge fields. 

That a  non-trivial solution does exist can be checked explicitly. We consider first the solution in which the scalar fields takes the real, diagonal form  
\beq   Q =   \left(\begin{array}{cc}\phi_1 & 0 \\0 & \phi_2\end{array}\right)\,.   \label{B=0sol}
\eeq
The scalar field $Q$  has precisely  this form in the known $B=0$  solution  in the theory without global $SU(2)_{R}$ symmetry  ($g_{R}=0;  \, A_{R \,i}\equiv 0$). 
Now in the presence of $SU(2)_R$ coupling,  the gauge fields will mix  (see Section~\ref{subwgauge}) and the vortex configuration will be modified. It is clear that 
in order to keep the tension unmodified, the right gauge fields $A_{R}$ must be turned on, to keep null the second term of Eq.~(\ref{eq:boghiggs}).  
An inspection of the BPS equations~(\ref{BPSeq}) however  shows that the solution for $g_{R}\ne 0$ is simply given by 
\beq     A^{a}_{R\, i}(x) =  \frac{g_{R}}{g_{L}}   \, A^{a}_{L\,  i}(x) \;,  \label{B=0gauge}
\eeq
the non-Abelian parts $A^{L}$ and $A^{R}$ in Eq.~(\ref{BPS2}) are identical. 
Once such this solution is found, all other color-``flavor''  rotated solutions (with $B \ne 0$)  can be constructed in the standard way by
\beq       Q \to  U\, Q \, U^{\dagger}, \qquad  A_{L\, i} \to   U\,  A_{L\, i}\, U^{\dagger}, \qquad A_{R\, i} \to   U\,  A_{R\, i}\, U^{\dagger}, \label{gaugetr}
\eeq
where $U\subset SU(2)$ is now a general reducing matrix, containing the coordinates of $\mathbb CP^{1}= SU(2)/U(1)$.  But in the presence of both left and right gauge fields~(\ref{gaugetr}) is
a genuine (global)  gauge transformation: the solutions generated are all gauge equivalent. 
Another way of stating this result is that the dependence   of the field configuration on the ``orientational modes'' $B$ can be gauged away.
This is nothing but the standard Higgs mechanism in which the Nambu-Goldstone modes are absorbed by the longitudinal components of the gauge fields, which become massive.

\subsection{Supersymmetry breaking}

 As well-known the BPS equations~(\ref{BPSeq}) amount to the conditions that a string-like soliton must  satisfy to preserve half of the supercharges of the original theory~\cite{Witten:1978mh,Edelstein:1993bb}. Even in weakly gauged case, therefore, non-Abelian vortices are 1/2 BPS, preserving four  residual supersymmetries. In general, we expect the low-energy physics for the zero-modes of 1/2 BPS solitons to be described by a  supersymmetric non-linear sigma model with four  ($\mathcal N=(2,2)$) supercharges. However, here we encounter an unexpected result: the effective theory we have derived is not supersymmetric, even though we have explicitly written only the bosonic part.
We thus have an intriguing  situation, which to the best of our knowledge has not been discussed before, where the low-energy physics in presence of a  BPS soliton is described by a non-supersymmetric effective theory.

Surprising though this result may appear, it has a simple physics explanation. As we have seen in the previous Section, since a vortex configuration breaks the residual $SU(2)_{V}$ gauge symmetry of the vacuum to $U(1)$, it  induces some effective mass terms for the non-diagonal fields $V_{\mu}^{\pm}$. It is easy to see that this ``vortex induced'' Higgs mechanism cannot be realized in a supersymmetric fashion. The Higgs mechanism is consistent with supersymmetry only when a massless vector multiplet can mix with a massless chiral multiplet to form a long massive vector multiplet. Another condition is that the chiral multiplet to be absorbed must contain only one NG boson (the other is called quasi-NG boson \cite{Kugo:1983ma}). In our case of $\mathbb CP^1$, scalar fields in $B$ are both NG bosons, so that Higgs mechanism is inconsistent with supersymmetry.

 In the  $SU(2)/U(1)$ breaking and consequent Higgs mechanism induced on the vortex, the two fields  $B$  and $B^{*}$  are precisely those absorbed by $V^{\pm}$.  In order for this phenomenon to occur supersymmetrically we would need two chiral superfields, 
but there is only one chiral multiplet $B$!\footnote{This point is a little similar to the fact that in supersymmetric standard model we need two Higgs doublet chiral superfields to give masses to the up and down fermions, whereas in the minimal Weinberg-Salam model  one complex doublet Higgs scalar is sufficient.}  Supersymmetry breaking is thus inevitable for the dynamics of the light fields $V_{\mu}$, in the presence of a vortex. Still, a static background with a vortex and vanishing vector gauge field $V_{\mu}$ preserves four  supercharges of the bulk theory.  

Indeed, before the right gauge fields are introduced, the  vortex effective action in supersymmetric theory is just the supersymmetric ${\mathbb C}P^{1}$ sigma model \cite{Novikov:1984ac, Morozov:1984ad,Higashijima:1999ki,Shifman:2004dr,Gudnason:2010rm},  
\beq
S_{1+1}^{\rm susy} = 2\beta \int dtdz\, d^{2}\theta\, d^{2}{\bar
{\theta}} \;
K(B, {\bar B}), \qquad  K= \tr\log\left(\mathbf{1} + B   {\bar B} \right)\;,
 \label{Susysigmamodel}
\eeq
where $B$ is now a chiral (${\bar B}$ antichiral)  superfield  containing the $\mathbb CP^{1}$  coordinate scalar  $B$. 
The unbroken color-flavor diagonal global symmetry $SU(2)$ symmetry is just the isometry group of the  ${\mathbb C}P^{1}$ sigma model.  Now we saw above that 
upon weak gauging of the flavor symmetry,  the diagonal vector-like $SU(2)$ group survives as an unbroken local gauge group, which clearly corresponds to the isometry group of ${\mathbb C}P^{1}$.   
The effective low-energy vortex action must then be 
the supersymmetric sigma model~(\ref{Susysigmamodel}) in which the full isometry group $SU(2)$ is gauged. 
As is well known  \cite{Bagger:1982fn} this cannot be achieved maintaining supersymmetry   for reasons already pointed out.   

With exactly  the same mechanism supersymmetry is broken in the non-Abelian vortex effective sigma model action  in all cases,  with $SU(N)$, $SO(2N)$, $USp(2N)$\cite{Gudnason:2010rm} gauge groups,  when a weak gauging of the flavor groups is introduced. 

\section{Discussion}

In this paper we have studied the effects of a weak gauging of the full $SU(N)_{F}$ flavor symmetry in an $U(N)$ gauge theory with $N_{F}=N$ massless scalar fields in the color-flavor locked phase, when a non-Abelian vortex is present. After the weak gauging, the color-flavor symmetry becomes an unbroken vector gauge symmetry with  massless $SU(N)$ gauge fields $V_{\mu}$, propagating in the bulk. As we have shown with Eq.~(\ref{eq:gaugedact}), the orientational degrees of freedom of the non-Abelian vortex then couple minimally to this gauge field. Since a generic vortex configuration breaks the $SU(N)_{V}$ gauge symmetry to  $U(N-1)_{V}$, the coupling of the gauge field $V_{\mu}$ to the non-Abelian vortex acts like an effective mass term for the gauge fields corresponding to the broken generators. We thus have a vortex-induced Higgs mechanism, where $2(N-1)$ massive vector fields eat up  the $N-1$ complex scalars $B$ describing the orientational modes of the vortex. The most dramatic effect of this mechanism is that the vortex excitations decay through a tunneling effect  to bulk excitations.

We have also considered the special circumstances when the model is embedded in an $\mathcal N=2$ supersymmetric theory. The weak gauging of the flavor symmetry can be done preserving full supersymmetry. The static vortex configuration preserves half of the supercharges and is BPS saturated, just as in the case with no weak gauging of the flavor symmetry.   However, the coupling of the massless gauge field $V_{\mu}$ to the non-Abelian vortex  necessarily breaks supersymmetry of  the worldsheet effective action. This (apparently) puzzling result, if consolidated by further checks,  is potentially a very interesting one with many possible applications,  and certainly deserves a more thorough study than is done here.  It might be thought that some subtle effects could restore a supersymmetric form of the quantum fluctuations along the vortex. 

Let us however summarize the main ingredients which entered our deduction. First we have shown that the weak gauging of the exact flavor symmetry leads to an unbroken diagonal (weak) gauge symmetry in the color-flavor locked vacuum. Second, the vortex effective action was shown to take, accordingly, the form of a gauged $\mathbb CP^{1}$ model. Thirdly, in an  ${\cal N}=2$ supersymmetric context, the latter becomes a ${\cal N}=(2,2)$
supersymmetric $\mathbb CP^{1}$ model, Eq.~(\ref{Susysigmamodel}), in which the full isometry group $SU(2)$ is weakly gauged. Supersymmetry is broken in such a setting because of lack of appropriate massless degrees of freedom.  Our argument thus relies upon a few simple basic points, and for that very reason seems to be rather robust. It might also be of some help to note that while the necessary condition for Coleman's theorem\cite{ColemanSymmetry}, i.e. the non degeneracy of the vacuum, is satisfied by our $4D$ bulk vacuum,  it is clearly  violated by the ``$2D$ vacuum'' (i.e., the minimum-energy vortex solution) immersed in it, where massless excitations are present. 
Our vortex effective action can be regarded as a low-energy realization of the Bagger-Witten model \cite{Bagger:1982fn}.

In all cases with  $SU(N)$, $SO(2n)$ and $USp(2N)$ gauge groups, the same economical structure of the effective $2D$ sigma model (with all massless scalars being NG bosons and no quasi NG bosons even in the supersymmetric version of the models) leads to spontaneous supersymmetry breaking on the vortex worldsheet.

It would also be very interesting to study more general situations. For example, one could gauge a $U(r)$ subgroup  of $SU(N)$ ($r<N-1$). The residual symmetry of the system would then be $U(r)\times SU(N-r)$, where $SU(N-r)$ would remain as a global symmetry.  The $\mathbbm CP^{N-1}$ moduli space of the original vortex is then reduced to  $U(r)\times SU(N-r)$ orbits. Vortices corresponding to the various strata will then have different tensions, and one has to determine which one corresponds to the vortices with the lowest tension. A similar situation has been already considered \cite{Vinci:2012mc} in the case of the semi-superfluid vortices~\cite{Balachandran:2005ev,Nakano:2007dr} present in the CFL phase of QCD~\cite{Alford:2007xm}. There, a $U(1)$ flavor symmetry is actually gauged, which corresponds to the electromagnetic interactions.   

In a supersymmetric context, where our vortex is BPS,   we can  consider the gauging of just the $U(1) \subset SU(2)$ global symmetry group  (``electromagnetic'' interactions), as in ~\cite{Alford:2007xm}.   In this case certain interesting features distinct from  those in the fully gauged case discussed in this paper will be present, such as the persistence of a superconducting current, similarly to the case of  Witten's  superconducting string~\cite{Witten:1984eb}. These will be discussed elsewhere.

\section*{Acknowledgments}

We thank Minoru Eto and Yutaka Ookouchi for discussions. 
This work of M.N. is supported in part by 
Grant-in Aid for Scientific Research (No.~23740198) 
and by the ``Topological Quantum Phenomena'' 
Grant-in Aid for Scientific Research 
on Innovative Areas (No.~23103515)  
from the Ministry of Education, Culture, Sports, Science and Technology 
(MEXT) of Japan. M.N.  thanks INFN, Pisa, for partial support and hospitality while this work was completed. 

\newpage	

\appendix
\section{Calculation of the effective action}
We first review the calculation of Ref.~\cite{Gudnason:2010rm}. 
It is much tidier to do the calculation with the form of the reducing matrix 
\begin{align}
U =
\begin{pmatrix}
\mathbbm{1} & - B^\dag \\
0 & \mathbbm{1}
\end{pmatrix}
\begin{pmatrix}
x^{-1} & 0 \\
0 & y^{-1}
\end{pmatrix}
\begin{pmatrix}
\mathbbm{1} & 0 \\
B & \mathbbm{1}
\end{pmatrix}
=
\begin{pmatrix}
x^{-1} & - B^\dag y^{-1} \\
B x^{-1} & y^{-1}
\end{pmatrix}
\label{eq:UmatrixBis} \ ,
\end{align}
with the matrices $x$ and $y$ defined by
\beq
x=  \sqrt{\mathbbm{1} + B^\dag B }  = X^{1/2}\ , \quad
y=  \sqrt{\mathbbm{1} + B B^\dag  }= Y^{1/2}\,
\eeq
rather than using $X$ and $Y$. 
This form, with $B$, $\mathbbm{1}$, $x$ and $y$  all $N\times N$ matrices,  is adequate for $SO(2N)$ and $USp(2N)$ models. For the $SU(N)$ model,  
$B$ is an
($N-1$)-component column-vector
\beq B =
\begin{pmatrix}
b_1 \\ \vdots \\ b_{N-1}
\end{pmatrix} \ , \label{eq:B_CPNvectorB} \eeq
while $B^{\dagger}$ is a corresponding row-vector;
\beq  x^{2}  =  X = 1 + B^\dag B \ , \qquad   y^{2}= Y = \mathbbm{1}_{N-1} + B B^\dag \ ,
\label{eq:XY_CPNA} \eeq
are a scalar and an $(N-1)\times (N-1)$ dimensional matrix, respectively, and the unit matrices  appearing in~(\ref{eq:UmatrixBis})  have appropriate ($1\times 1$ and $N-1 \times N-1$) dimensions.

By repeatedly making use of relations like
\beq     x^{-1} \, B^{\dagger} =   B^{\dagger}\, y^{-1}, \qquad   y^{-1} \, B = B \, x^{-1}, 
\eeq
one finds easily
\beq   U^{-1} \de_{\alpha} U =  \left(\begin{array}{cc}x^{-1} B^{\dagger} \de_{\alpha} B \, x^{-1} - \de_{\alpha} x\, x^{-1} & -x^{-1} \de_{\alpha} B^{\dagger} y^{-1} \\y^{-1} \de_{\alpha} B x^{-1} & y^{-1} \de_{\alpha} B^{\dagger}  y^{-1} -  \de_{\alpha} y \, y^{-1}\end{array}\right)\,.
\eeq
Having defined
\beq    T=  \left(\begin{array}{cc}\mathbbm{1} & 0 \\0 & -\mathbbm{1}\end{array}\right)
\eeq 
with appropriate dimensions for the unit matrices,
one gets 
\beq    (U^{-1} \de_{\alpha} U)_{\perp}= \frac{1}{2} [\, U^{-1} \de_{\alpha} U - T \, U^{-1} \de_{\alpha} U \, T \,] =  
\left(\begin{array}{cc}0  & -x^{-1} \de_{\alpha} B^{\dagger} y^{-1} \\ y^{-1} \de_{\alpha} B x^{-1} & 0  \end{array}\right)
\eeq
as in Ref.~\cite{Gudnason:2010rm}.  The usual form of the effective sigma model action easily follows then from $\Tr  (U^{-1} \de_{\alpha} U)_{\perp}^{2}$.

Now, as $U$ transforms linearly under the unbroken gauge symmetry,  the replacement is
\beq   U^{-1} \de_{\alpha} U  \to    U^{-1} \nabla_{\alpha} U, \qquad  \nabla_{\alpha} U  = (\de_{\alpha} +  i g_{V}   V_{\alpha})   U.
\eeq
 Let us decompose the vector gauge fields into a block-wise form: 
\beq    V_{\alpha}=   \left(\begin{array}{cc}v_{\alpha} & w_{\alpha} \\  w^{\dagger}_{\alpha} & z_{\alpha}\end{array}\right), \qquad  \Tr\, V_{\alpha} =0, 
\label{eq:dec}
\eeq
then
\beq    (U^{-1} V_{\alpha} U )_{\perp} =  
\left(\begin{array}{cc}0  & -x^{-1}  {\tilde w}_{\alpha} y^{-1} \\ y^{-1}  {\tilde w}_{\alpha}^{\dagger}  x^{-1} & 0  \end{array}\right),\nonumber \\
\nonumber \\
{\tilde w}_{\alpha}\equiv   w_{\alpha} - v_{\alpha} B^{\dagger} +  B^{\dagger} z_{\alpha} -  B^{\dagger} w_{\alpha} B^{\dagger}.
\eeq
Therefore
\beq    (U^{-1} \nabla_{\alpha} U )_{\perp} =   \left(\begin{array}{cc}0  & -x^{-1}( \hat{ \nabla}_{\alpha} B )^{\dagger} y^{-1} \\ y^{-1} \hat{\nabla}_{\alpha} B x^{-1} & 0  \end{array}\right),
\eeq
where
\beq    \hat{\nabla}_{\alpha} B  \equiv \de_{\alpha}  B  + i g ( z_{\alpha} B - B v_{\alpha}  + w_{\alpha}^{\dagger} -   B w_{\alpha} B )\;,
\label{defined1}\eeq
\beq    (\hat{\nabla}_{\alpha} B)^{\dagger}  =   \de_{\alpha}  B^{\dagger}  -   i g ( B^{\dagger} z_{\alpha} - v_{\alpha} B^{\dagger}  + w_{\alpha} -  B^{\dagger} w_{\alpha}^{\dagger} B^{\dagger} ) \,. \label{defined2}
\eeq
Notice the non-linear form of the covariant derivative acting on the fields $B$. As done in the text for the $SU(2)$ case, we can identify massive and massless combinations of $V$ using the following change of coordinates:
\begin{equation}
V_{\alpha}'\equiv U^{-1} V_{\alpha}  U
\end{equation}
in terms of which:
\beq    (U^{-1} V_{\alpha} U )_{\perp} =  (V^{'}_{\alpha})_{\perp}=
\left(\begin{array}{cc}0  & \omega_{\alpha}'\\
 {\omega_{\alpha}^{\dagger}}' & 0  \end{array}\right)\,,
\eeq
The massive combination ${\omega_{\alpha}}'$ then enter the covariant derivative of $B$ as follows
\begin{equation}
 \hat{\nabla}_{\alpha} B  \equiv \de_{\alpha}  B  + i g_{V} x \omega_{\alpha}' y\;.
\end{equation}

\subsection*{Calculation of the effective action in terms of $n$}
\label{sec:Bnconn}
The relation between the inhomogeneous and homogeneous coordinates $B$ and $n$ can be written in the following way~\cite{Shifman:2004dr,Shifman:2007ce}:
\begin{eqnarray}
 n  = 
\begin{pmatrix}
X^{-\frac{1}{2}} \\
B X^{-\frac{1}{2}}
\end{pmatrix}=U\, n_{0},\quad n_{0} =  
\left( \begin{array}{c}
  1 \\
   0    
\end{array}
\right)  \ . \label{identi}
\end{eqnarray}
The relations above are not enough to determine $U$ in terms of $n$ completely, but this is not needed, as we shall see. In terms of $n$ we can in fact write the following:
\begin{eqnarray}
\frac12\, U \,T \,U^{-1}=-n\,n^{*}+\frac12&\,,  \nonumber \\
\end{eqnarray}
from which we have, unambiguously:
\beq    (U^{-1} \de_{\alpha} U)_{\perp}=  i \bigg(   \partial _{\alpha} n \,n^{*}-n \,\partial_{k}n^{*}- 2n\,n^{*}(n^{*}\cdot \partial_{\alpha}n)   \bigg)\,.
\eeq
From the definition of $n$ we see that it transforms linearly like $U$ under the unbroken gauge symmetry. The substitution in the gauged case is then analogous to the previous section:
\begin{equation}
\de_{\alpha} n \rightarrow \nabla_{\alpha} n  = (\de_{\alpha} +  i g_{V}   V_{\alpha})   n\,.
\end{equation}
From the relations above we determine the form of the kinetic terms in the effective action in terms of $n$
\begin{equation}
\Tr \, (U^{-1}\nabla_{\alpha} U)_{\perp}^{2}=-\left( \nabla_{\alpha}n^{*}\cdot\nabla_{\alpha}n+(\nabla_{\alpha}n^{*}\cdot n)^{2}\right)\,.
\end{equation}

\newpage

\bibliography{Bibliographysmart}
\bibliographystyle{nb}

\end{document}